\documentclass[a4paper, 11pt]{article}
\usepackage{jheppub}		
\usepackage{multirow} 
\usepackage{graphicx}
\usepackage{pdfpages}
\usepackage{dcolumn}
\usepackage{bm}
\usepackage{placeins} 
\usepackage{hyperref} 
\usepackage{appendix} 
\usepackage{amssymb,amsmath} 
\usepackage{amsthm}
\usepackage{mathtools}
\usepackage{epstopdf}
\usepackage{verbatim}
\usepackage{youngtab} 
\usepackage{float}
\usepackage{makecell}
\usepackage{ bbold }
\usepackage{diagbox}
\DeclareMathOperator{\mm}{\mathcal{M}}

\title{$N = 3$ SCFTs in 4 dimensions and non-simply laced groups}
\author{Mikhail Evtikhiev
\\
\it{Department of Particle Physics and Astrophysics,\\
Weizmann Institute of Science, Rehovot 7610001, Israel}
}
\emailAdd{Mikhail.Evtikhiev@weizmann.ac.il}
\abstract{
	In this paper we discuss various $N=3$ SCFTs in 4 dimensions and in particular those which can be obtained as a discrete gauging of an $N=4$ SYM theories with non-simply laced groups. The main goal of the project was to compute the Coulomb branch superconformal index and Higgs branch Hilbert series for the $N=3$ SCFTs that are obtained from gauging a discrete subgroup of the global symmetry group of $N=4$ Super Yang-Mills theory. The discrete subgroup contains elements of both $SU(4)$ R-symmetry group and the S-duality group of $N=4$ SYM. This computation was done for the simply laced groups (where the S-duality groups is $SL(2, \mathbb{Z})$ and Langlands dual of the the algebra $\prescript{L}{}{\mathfrak{g}}$ is simply $\mathfrak{g}$) by Bourton et al. \cite{Bourton}, and we extended it to the non-simply laced groups. We also considered the orbifolding groups of the Coulomb branch for the cases when Coulomb branch is relatively simple; in particular, we compared them with the results of Argyres et al. \cite{Argyres1}, who classified all $N\geq 3$ moduli space orbifold geometries at rank 2 and with the results of Bonetti et al. \cite{Bonetti}, who listed all possible orbifolding groups for the freely generated Coulomb branches of $N\geq 3$ SCFTs. Finally, we have considered sporadic complex crystallographic reflection groups with rank greater than 2 and analyzed, which of them can correspond to an $N=3$ SCFT.
}
\begin{document}

\maketitle

\section{Introduction}
In the last few years there have been many advances in the studies of superconformal field theories with extended supersymmetry in 4 dimensions. A particularly fruitful area of research was research of $N=3$ SCFTs; for example, in \cite{Aharony} it was shown, that the relation between dimensions of Coulomb branch operators and $2a-c$ \cite{Shapere} is only true if theory doesn't possess a discrete gauge group, and in \cite{Bourton} it was shown, that, contrary to a long-standing belief, Coulomb branch doesn't have to be freely generated or even to be a complete intersection manifold and, in fact, is not a complete intersection manifold for a ``generic'' $N=3$ SCFT. 

Large progress has also been made in classifying superconformal field theories. In the series of papers \cite{Argyres1, Argyres2, Argyres3, Argyres4, Argyres5} by Argyres et al. authors have studied the Coulomb branches of $N\geq 2$ SCFTs and have employed various methods to probe the relations in the holomorphic polynomial ring and the orbifolding structure of the Coulomb branch manifold. In particular, in \cite{Argyres2} Argyres and Martone have suggested a method to refine the Coulomb limit of the superconformal index, that simplifies tracking the relations in the coordinate ring. In \cite{Argyres1} they have classified Coulomb branches for rank-2 SCFTS with $N>2$ supersymmetry. It is also possible to study Coulomb branch manifold in a ``bottom-up'' approach; in \cite{Bonetti} Bonetti et al. have considered a class of $N = 2$ vertex operator algebras $W_G$ labeled by crystallographic complex reflection groups, that are extensions of the $N = 2$ super Virasoro algebra obtained by introducing additional generators; they have also found a way to recover the Macdonald limit of the superconformal index of the parent $4d$ theory from the corresponding vertex operator algebra, when such a theory exists. Their construction is also interesting because for every rank-$r$ $N>2$ SCFT with a freely generated Coulomb branch its complex structure can be written as $\mathbb{C}^r /\Gamma$, where $\Gamma$ is a crystallographic complex reflection group acting irreducibly on $\mathbb{C}^r$. However, not every crystallographic complex reflection group $\Gamma$ will correspond to some $N=3$ SCFT. For example, none of the rank-2 crystallographic complex reflection groups $G_4$, $G_5$, $G_8$\footnote{Notation for the complex crystallographic reflection groups is in agreement with with Shephard and Todd \cite{Todd}.} can be an orbifolding group for $N>2$ SCFT \cite{Argyres1, Caorsi}.

Another way to analyze the landscape of superconformal field theories is by constructing them and then computing their index. A particular example of such study can be found in a paper by Bourton et al. \cite{Bourton}, where authors have computed Coulomb limit of the superconformal index and Higgs branch Hilbert series for various $N=3$ and $N=4$ SCFTs and analyzed the Coulomb branches of these theories. To construct new theories, they have considered $N=4$ SCFTs with simply laced gauge groups and noticed, that these theories have an enhanced discrete global symmetry at certain values of the gauge coupling. They then refined obtained superconformal index by a fugacity for the enhanced discrete symmetry. The index of the discretely gauged daughter theory is then obtained by ``integrating'' over the additional fugacity, which takes values in the discrete group. The enhanced discrete symmetry is constructed from a subgroup of S-duality group\footnote{It is important to notice that analysis of Bourton et al. doesn't account for the line operators, so the discrete global symmetry isn't always present and not every $N=3$ theory they list actually exists \cite{SeibergLines}, see appendix \ref{sec:bourton} for more details.}, so authors restrict their studies to the case of simply laced gauge groups, leaving the non-simply laced groups for the further studies.

In this paper we continued studying the landscape of $N>2$ SCFTs in 4 dimensions, found more new $N=3$ theories in the spirit of \cite{Bourton} and bridged some of the gaps between the research of Argyres et al. \cite{Argyres1, Argyres2}, Bonetti et al. \cite{Bonetti}, and Bourton et al. \cite{Bourton}. In order to do that, we extended the results of \cite{Bourton} to the non-simply laced groups. We also analyzed the geometry of the moduli space of the theories we obtained; for the cases when the Coulomb branch is freely generated the orbifolding group is a complex reflection group in agreement with \cite{Bonetti}. The Coulomb branches of the rank-2 theories we found are in agreement with the results of \cite{Argyres1}. Finally, we considered sporadic crystallographic complex reflection groups\footnote{The non-sporadic groups were considered in \cite{Caorsi}.} and, using the methods of \cite{Caorsi}, checked, which of them may be an orbifolding group for $N=3$ SCFTs; we compute Higgs branch Hilbert series for $N=3$ SCFTs that can originate from these groups.

\section{\texorpdfstring{$N=3$}{N=3} SCFTs from gauging \texorpdfstring{$N=4$}{N=4} SCFTs with non-simply laced gauge groups}

\subsection{Coulomb branch index computation} \label{CBO}

The first part of the computation is to compute Coulomb branch limit index. It is more instructing to do these computations along the lines of \cite{Argyres2}, since the results are similar to the computations done as in \cite{Bourton}, but the method from \cite{Argyres2} gives more refined version of the index that simplifies analysis of the Coulomb branch manifold. The computation method is as follows:
\begin{itemize}
	\item[1.] We start with an $N=4$ SYM theory with a non-simply laced gauge group ($B_n, C_n, F_4$ or $G_2$); the simply laced cases have been discussed in \cite{Bourton}. The non-simply laced case is more complicated than the simply laced one, so writing the S-duality group and finding the discrete symmetry group is slightly trickier. For the simply laced groups the S-duality group is simply $SL(2,\mathbb{Z})$, while for the non-simply laced groups it is rather Hecke group $\Gamma_0(q) \equiv H_{2q}$, where $q$ is the square of length ratio of the short roots to the long ones; $q = 2$ for $B_n, C_n, F_4$ and $q=3$ for $G_2$. $SL(2,\mathbb{Z})$ is generated by the three elements $S = \left(\begin{smallmatrix}0 & -1\\ 1 & 0\end{smallmatrix}\right)$, $T = \left(\begin{smallmatrix}1 & 1\\ 0 & 1\end{smallmatrix}\right)$, $C = -1$, while $\Gamma_0(q)$ is generated by $C$, $T$, $A = ST^q S$. $SL(2,\mathbb{Z})$ generators obey the relations $C^2 = 1$, $S^2 = (ST)^3 = C$, while for $\Gamma_0(q)$ $(AT)^q = C$. It is important to notice that for almost every algebra (with the exception of $A_n$) $C$ is a part of Weyl group and so acts on the theory trivially; only the quotient of the S-duality group by its center acts faithfully on the theory. For $B_n$, $C_n$ it is enough to consider $AT$, while $G_2$, $F_4$ require more attention. As $B_n$, $C_n$ theories have the same Coulomb and Higgs branches, superconformal indices we consider in this paper cannot distinguish between these theories, so we will restrict ourselves to considering $B_n$ theories from now on.
	\item[2.] For $G_2$ and $F_4$ similarly to \cite{AKS} we should rather use a transformation $\tilde{S}$ such that $\tilde{S} T \tilde{S} = ST^qS$ so our S-duality group will look more like $SL(2,\mathbb{Z})$, as $C, T, \tilde{S}$ will obey the relations $C^2 = 1$, $\tilde{S}^2 = (\tilde{S} T)^{2q} = C$. This is because $\tilde{S}$ takes us back to the same group and if we will consider only $AT$ transformations similarly to $B_n$ case, we will miss some of the discrete symmetries. $\tilde{S}$ is not a part of the Weyl group, so it acts on the moduli space non-trivially; in \cite{AKS} it was shown that for $G_2$ the Coulomb branch operators transform as $(U_2, U_6) \xrightarrow{\tilde{S}} (U_2, -U_6)$, and for $F_4$ the rule is $(U_2, U_6, U_8, U_{12}) \xrightarrow{\tilde{S}} (U_2, -U_6, U_8, -U_{12})$.
	\item[3.] Now we can find, for which values of the coupling $\tau$ the discrete subgroups of the S-duality group for the various theories we have considered above will leave $\tau$ unchanged. For $G_2$ we have three different options \cite{AKS}: we can consider either $\tau = \frac{i}{\sqrt{3}}$, which is fixed under $\mathbb{Z}_2 \times \mathbb{Z}_2^c$, where $\mathbb{Z}_2$ is generated by $\tilde{S}$ and $\mathbb{Z}_2^c$ is generated by $C$ or $\tau = -\frac12 \pm \frac{i}{2\sqrt{3}}$\footnote{Apparently there is a typo in \cite{AKS}.}, for which we have two options for the symmetry group: $\mathbb{Z}_6 \times \mathbb{Z}_2^c$ ($\mathbb{Z}_6$ is generated by $(\tilde{S}T)$), and $\mathbb{Z}_3  \times \mathbb{Z}_2^c \subset \mathbb{Z}_6  \times \mathbb{Z}_2^c$ discrete symmetry with $\mathbb{Z}_3$ generated by $(\tilde{S}T)^2$. For $F_4$ we also have three different options: we have enhanced $\mathbb{Z}_2 \times \mathbb{Z}_2^c$ symmetry at $\tau = \frac{i}{\sqrt{2}}$ (generated by $\tilde{S}$), $\mathbb{Z}_4 \times \mathbb{Z}_2^c$ symmetry at $\tau = -\frac12 \pm \frac{i}{2}$ (generated by $(\tilde{S}T)$), and a $\mathbb{Z}_2  \times \mathbb{Z}_2^c \subset \mathbb{Z}_4  \times \mathbb{Z}_2^c$ discrete symmetry with the generator $(\tilde{S}T)^2$, it is realized differently comparing to the $\mathbb{Z}_2$ generated by $\tilde{S}$. For $B_n$ groups we only have $\mathbb{Z}_2 \times \mathbb{Z}_2^c$ at $\tau = -\frac12 \pm \frac{i}{2}$ (generated by $(AT)$). As $C$ belongs to the Weyl group and is therefore a trivial operation, only the quotient of the Hecke subgroup by its center acts faithfully on the $N=4$ theory, so we drop the $\mathbb{Z}_2^c$ from now on.
	\item[4.] The S-duality transformations transform the chiral supercharges by a phase \cite{Kapustin}. Namely, if an element of the S-duality group transforms the SYM coupling as
	\begin{equation}
		\sigma = \begin{pmatrix} a & b \\ c & d\end{pmatrix}: \qquad \tau \rightarrow \frac{a\tau + b}{c\tau + d},
	\end{equation}
	then the chiral supercharges transform as
	\begin{equation}
		Q^i_\alpha \rightarrow e^{i\chi} Q^i_\alpha; \qquad e^{i\chi} = \left(\frac{|c\tau + d|}{c\tau + d}\right)^{1/2}.
	\end{equation} 
	In particular, the transformations we have found above transform the chiral supercharges as following:
	\begin{center}
		\begin{tabular}{|c|c|c|c|} \hline
 			\diagbox{$\mathfrak{g}$}{$\sigma$} & $\tilde{S}$ & $\tilde{S}T$ & $(\tilde{S}T)^2 = AT$ \\ \hline
 			$G_2$ & $Q^i_\alpha \rightarrow e^{-i\pi/4} Q^i_\alpha$ & $Q^i_\alpha \rightarrow e^{-i\pi/12} Q^i_\alpha$ & $Q^i_\alpha \rightarrow e^{-i\pi/6} Q^i_\alpha$ \\ \hline
 			$F_4$ & $Q^i_\alpha \rightarrow e^{-i\pi/4} Q^i_\alpha$ & $Q^i_\alpha \rightarrow e^{-i\pi/8} Q^i_\alpha$ & $Q^i_\alpha \rightarrow e^{-i\pi/4} Q^i_\alpha$\\ \hline \label{tftable}
 			$B_n$ & --- & --- & $Q^i_\alpha \rightarrow e^{-i\pi/4} Q^i_\alpha$\\ \hline
		\end{tabular}
	\end{center}
 	Now we need to offset the action of the S-duality transformation; to do that, we will use elements of the R-symmetry group $SU(4)_R = SO(6)_R$. We can organize six real adjoint scalar fields $\phi^I, I \in \mathbf{6}$ of $SU(4)_R$ into a triplet of complex scalars $\varphi^a, a \in \mathbf{3}$ of $U(3)$; $\varphi^a = \phi^{2a-1}+i\phi^{2a}$. The R-symmetry group element $\rho$ can be represented by a simultaneous rotation in three orthogonal planes in $\mathbb{R}^6 \simeq \mathbb{C}^3$:
 	\begin{equation}
 		\rho = \begin{pmatrix} e^{i\psi_1}& & \\ & e^{i\psi_2}& \\ & & e^{i\psi_3}\end{pmatrix} \in U(3) \subset SU(4)_R.
 	\end{equation}
 	Then $\rho$ rotates the complex scalars by a phase $\varphi_a \xrightarrow{\rho} e^{i\psi_a} \varphi_a$, and the four chiral supercharges transform as 
 	\begin{align}
 		& Q^1_\alpha \xrightarrow{\rho} e^{i(\psi_1+\psi_2+\psi_3)/2} Q^1_\alpha \\
 		& Q^2_\alpha \xrightarrow{\rho} e^{i(\psi_1-\psi_2-\psi_3)/2} Q^2_\alpha \\
 		& Q^3_\alpha \xrightarrow{\rho} e^{i(-\psi_1+\psi_2-\psi_3)/2} Q^3_\alpha \\
 		& Q^4_\alpha \xrightarrow{\rho} e^{i(-\psi_1+\psi_2+\psi_3)/2} Q^4_\alpha \label{chargetransform}
 	\end{align}
 	Now we can choose $\psi_a$ in such a fashion that the combination of $\rho, \sigma$ will leave $Q^1, Q^2, Q^3$ invariant; this means that for a given $\sigma$, $\rho$ should be equal to $\rho = \mathrm{diag}(e^{2i\pi/n}, e^{2i\pi/n}, e^{-2i\pi/n})$, where $n$ is the value of the denominator in the exponent at the corresponding cell of the \ref{tftable}. All in all, resulting Coulomb branch will be described by $\mathbb{C}^r / (\Gamma \rtimes \Gamma_k)$, where $\Gamma_k$ can be $\mathbb{Z}_2, \mathbb{Z}_3, \mathbb{Z}_6$ for $G_2$, $\mathbb{Z}_2, \mathbb{Z}_2'$ (different from $\mathbb{Z}_2$), and $\mathbb{Z}_4$ for $F_4$, and $\mathbb{Z}_2$ for $B_n$ theories, with $\Gamma$ denoting the orbifolding group of the original theory.
	\item[5.] Using this knowledge, we can now compute the refined Molien series as in \cite{Argyres2} (see equation (4.13) and derivation around it):
	\begin{equation}
		P_{J_{\Gamma_k}}(t_1, \ldots, t_l) = \frac{1}{|\Gamma_k|}\sum\limits_{g\in \Gamma_k}\frac{1}{\det \left(1 - g\; \mathrm{diag}(t_1,\ldots, t_l)\right)} \label{Molien},
	\end{equation}
	where $t_i$ are coordinates that correspond to the Coulomb branch operators of the original theory. Applying the plethystic logarithm\footnote{$\mathrm{PLog}$ is plethystic logarithm: $\mathrm{PLog}(f(t)) = \sum\limits_{n=1}^\infty \frac{\mu(m)}{m} \log f(t^m)$, where $\mu(m)$ is the M\"obius function.} to $P_{J_{\Gamma_k}}$ one obtains the generators of the Coulomb branch of the resulting theory, as well as the relations between them in the form
	\begin{equation}
		\mathcal{F}_\Gamma(t) = \sum\limits_k c_k^+ t^k - \sum\limits_{k'} c^-_{k'} t^{k'},
	\end{equation}
	where the positive coefficients count the number of generators of degree $k$ and the negative ones count the number of relations at degree $k'$. If the Coulomb branch turns out to be not a complete intersection manifold, then $\mathcal{F}_\Gamma(t)$ should be not a polynomial, but rather an infinite power series\footnote{This is not always true: there can be “unexpected” cancellations between factors in the numerator and denominator of the Molien series series \eqref{Molien} \cite{Argyres2}. This can happen when the degree of a relation happens to be the same as that of an affine parameter in the coordinate ring, or if the degree of a syzygy happens to coincide with that of a relation, etc. As the Coulomb branch rank increases, such accidental cancellations become more likely, but, at least for the low-rank examples, one might expect that the plethystic logarithm will accurately capture the degrees and counting of generators and relations.}.
	\item[6.] We can now do a cross-check of the results we have obtained by considering the orbifolding group of the Coulomb branch directly. The orbifolding group $\Gamma$ of the Coulomb branch in the original $N=4$ theory is a Weyl group of a Lie algebra and, therefore, a crystallographic Coxeter group. If the Coulomb branch is freely generated, $\Gamma \rtimes \Gamma_k$ should be a crystallographic complex reflection group (see \cite{Bonetti} Table 1 for the table of the irreducible crystallographic complex reflection groups, divided into non-Coxeter and Coxeter groups; from it one can also read the dimensions of the Coulomb branch generators). Thus the cross-check is done by computing $\Gamma \rtimes \Gamma_k$ and checking, if the resulting group is a crystallographic complex reflection group; if it is, then the degrees of its fundamental invariants should match the dimensions of the Coulomb branch operators obtained from the Molien series computations results.
\end{itemize}

\subsection{Higgs branch Hilbert series computation}
Next we can move on to the computation of Higgs branch Hilbert series. The algorithm for the computation is as follows
\begin{itemize}
	\item[1.] The Higgs branch Hilbert series can be constructed in a similar fashion. We can use the fact that $N=2$ Higgs branch and $N=2$ Coulomb branch are parts of the moduli space that is defined as 
	\begin{equation}
		\mm_\Gamma = \mathbb{C}^{3r} / \rho_\tau(\Gamma) = \mathbb{C}^{3r} / (\mathbf{1}_3 \otimes \mu_\tau(\Gamma)).
	\end{equation}
	The complex structure of $\mm_\Gamma$ is determined by picking one left-handed supercharge in the $N = 3$ algebra and calling the complex scalars which are taken to left-handed Weyl spinors by the action of that supercharge the holomorphic coordinates on $\mm_\Gamma$. The special coordinates on $\mm_\Gamma$ are not holomorphic; from every $SU(3)_R$ triplet two can be taken to be holomorphic and the third anti-holomorphic. Thus, for example,
	\begin{equation}
		(z^1_i, z^2_i, z_{3i}) \equiv (a_i^1, a_i^2, \bar{a}_{3i}); 1 \leq i \leq r,
	\end{equation}
	can be taken as the holomorphic coordinates (see discussion near eq. \eqref{chargetransform}). When we choose a $N=2$ subalgebra of $N=3$, we choose a minimally embedded $SU(2)_R \subset SU(3)_R$. Then the subspace fixed by the $SU(2)_R$ is the $N=2$ Coulomb branch. If we now assume that $\mm_\Gamma$ is an orbifold, and $\mu_\tau: \Gamma \rightarrow GL(r, \mathbb{C})$ then it can be written as
	\begin{equation}
		\mm_\Gamma \equiv \mathbb{C}^{3r} / \mu_\tau (\Gamma) \oplus \mu_\tau (\Gamma) \oplus \bar{\mu_\tau} (\Gamma),
	\end{equation}
	with the Coulomb branch $\mathcal{C}_\Gamma \equiv \mathbb{C}^{r} / \mu_\tau (\Gamma)$ and Higgs branch 
	\begin{equation}
		\mathcal{H}_\Gamma \equiv \mathbb{C}^{2r} / \mu_\tau (\Gamma) \oplus \bar{\mu_\tau} (\Gamma).
	\end{equation}
	Therefore, we can construct the Higgs branch Hilbert series in a fashion similar to the Coulomb branch (see \cite{Argyres2} for more details).
	\item[2.] The usual Higgs branch Hilbert series (the unrefined version) has only one fugacity that tracks scaling dimensions of the operators. Since $\mm_\Gamma$ carries a non-holomorphic $U(3)_R$ isometry, we can refine the Hilbert series as
	\begin{align}
		&H_{\mm_\Gamma} (t, v, u_1, u_2) = \\
		&=\frac{1}{|\Gamma|} \sum\limits_{g\in \Gamma} \frac{1}{\det (\mathbf{1} - tvu_1 \mu_\tau(g))} \frac{1}{\det (\mathbf{1} - tv\frac{u_2}{u_1} \mu_\tau(g))} \frac{1}{\det (\mathbf{1} - t\frac{u_2}{v} \overline{\mu_\tau}(g))} \label{bigseries}
	\end{align}
	The fact that the Hilbert series factorizes in three pieces is an immediate consequence of the fact that the group action on $\mathbb{C}^{3r}$ is chosen to be a direct sum of three factors $\rho = \mu_\tau \oplus \mu_\tau \oplus \overline{\mu_\tau}$, each of which acts independently on $\mathbb{C}^r$. The choice of fugacities is in agreement with the $U(3)_R$ weights of the holomorphic coordinates ($[1; 0]_1$ for $z_i^1$, $[-1; 1]_1$ for $z_i^2$, $[0; 1]_{-1}$ for $\bar{z}_i^3$); $u_1, u_2$ fugacities powers are the $SU(3)_R$ weights, $t$ corresponds to the scaling dimension and $v$ tracks the $U(1)_R$ charge. We can now reduce \eqref{bigseries} to obtain Molien formula for Higgs and Coulomb branches: 
	\begin{align}
		&H_{\mathcal{C}_\Gamma} (t, v, u_1, u_2) = \frac{1}{|\Gamma|} \sum\limits_{g\in \Gamma} \frac{1}{\det (\mathbf{1} - tvu_1 \mu_\tau(g))} \label{CBG}\\
		&H_{\mathcal{H}_\Gamma} (t, v, u_1, u_2) = \frac{1}{|\Gamma|} \sum\limits_{g\in \Gamma}  \frac{1}{\det (\mathbf{1} - tv\frac{u_2}{u_1} \mu_\tau(g))} \frac{1}{\det (\mathbf{1} - t\frac{u_2}{v} \overline{\mu_\tau}(g))} \label{HBG}
	\end{align}
	The definition of the Higgs branch Hilbert series according to \cite{Argyres1} takes the whole series \eqref{bigseries}, while in \cite{Bourton} the authors restrict to the smaller series \eqref{HBG}; this can be seen by comparing eqn. (6.14) of \cite{Bourton} to eqn. (4.10) of \cite{Argyres1}. We will stick to the definition chosen in \cite{Argyres1}, since it is formulated in the $N=3$ language.
	\item[3.] Now let us find how $N=3$ multiplets contribute to the Higgs branch Hilbert series. In order to do that, one has to find the embedding of the group $\Gamma$ in $GL(r, \mathbb{C})$. In our case $\Gamma = \mathcal{W}(\mathfrak{g}) \rtimes \mathbb{Z}_n$, so we only need to find how to embed $\mathbb{Z}_n$ properly. To check whether the $\mathbb{Z}_n$ embedding we've chosen is correct we can plug in $\Gamma$ into \eqref{CBG} and compare the results with the ones we obtained with the other method in subsection \ref{CBO}. One more sanity check is to consider terms up to $t^3$ in the \eqref{bigseries} expansion; according to \cite{Argyres1} the expansion for $N=3$ SCFTs should go as 
	\begin{equation}
		\mathcal{I}_H = t^2(u_1 u_2 + \frac{u_2^2}{u_1}) + O(t^3).
	\end{equation}
\end{itemize}

\subsection{Results}
\subsubsection{Gauging \texorpdfstring{$N=4$ $G_2$}{N=4 G2} SYM}
The action of the generators $\mathcal{C}_l$ of the $\mathbb{Z}_l$ groups on the Coulomb branch coordinates is
\begin{equation}
	\mathcal{C}_2 = \begin{pmatrix} -1 & 0 \\ 0 & 1 \end{pmatrix}, \qquad \mathcal{C}_3 = \begin{pmatrix} \exp(\frac{4i\pi}{3}) & 0 \\ 0 & 1 \end{pmatrix}, \qquad \mathcal{C}_6 = \begin{pmatrix} \exp(\frac{2i\pi}{3}) & 0 \\ 0 & -1 \end{pmatrix},
\end{equation}
and the dimensions of the Coulomb branch operators are $\Delta = 4, 6$ for $\Gamma_2$ gauging of the theory, $\Delta = 6, 6$ for $\Gamma_3$ gauging and $\Delta = 6, 12$ for $\Gamma_6$ gauging. In every case Coulomb branch is freely generated.

The orbifolding group for the original manifold is $\mathrm{Weyl}(\mathfrak{g}_2) = G(6, 6, 2)$ (we use Shephard and Todd notation for the complex reflection groups), and it is easy to check directly that $G(6, 6, 2) \rtimes \mathbb{Z}_2 = G(6, 3, 2)$, $G(6, 6, 2) \rtimes \mathbb{Z}_3 = G(6, 2, 2)$, $G(6, 6, 2) \rtimes \mathbb{Z}_6 = G(6, 1, 2)$. These groups are also present in the Table 1 of \cite{Argyres1}\footnote{$G(6, 2, 2)$ is written in \cite{Argyres1} as $\mathrm{Weyl}(\mathfrak{su}_3) \rtimes \mathbb{Z}_6$.}, so our result match theirs and fill some of the gaps in the classification of the $N\geq 3$ SCFTs with rank-2 moduli spaces.

The Higgs branch Hilbert series for $\mathbb{Z}_2$ gauging is 
\begin{align}
	&t^2 \left(\frac{u_2^2}{u_1}+u_1 u_2\right)+t^4 \left(u_1^4 v^4+\frac{u_2^4 v^4}{u_1^4}+\frac{u_2^3 v^4}{u_1^2}+\right.\\
	&\left.+2 u_2^2 v^4+u_1^2 u_2 v^4+\frac{u_2^4}{v^4}+\frac{u_2^4}{u_1^2}+u_2^3+u_1^2 u_2^2\right)+ O(t^6),
\end{align}
for $\mathbb{Z}_3$ gauging it is given by
\begin{equation}
	t^2 \left(\frac{u_2^2}{u_1}+u_1 u_2\right)+ t^4 \left(\frac{u_2^4}{u_1^2}+u_2^3+u_1^2 u_2^2\right)+ O(t^6)
\end{equation}
and for $\mathbb{Z}_6$ gauging it is given by
\begin{equation}
	t^2 \left(\frac{u_2^2}{u_1}+u_1 u_2\right)+ t^4 \left(\frac{u_2^4}{u_1^2}+u_2^3+u_1^2 u_2^2\right)+ O(t^6);
\end{equation}
we can use $\mathcal{C}_2$, $\mathcal{C}_3$, $\mathcal{C}_6$ as the generators of $\mathbb{Z}_n$; indices for $\mathbb{Z}_3$, $\mathbb{Z}_6$ gaugings differ at the $t^6$ order.

\subsubsection{Gauging \texorpdfstring{$N=4$ $F_4$}{N=4 F4} SYM}
The action of the generators $\mathcal{C}_l$ of the $\mathbb{Z}_l$ groups on the Coulomb branch coordinates is
\begin{equation}
	\mathcal{C}_2 = \mathrm{diag}(-1, 1, 1, -1), \qquad \mathcal{C}_4 = \mathrm{diag}(i, i, 1, 1), \qquad \mathcal{C}_2' = \mathrm{diag}(-1, -1, 1, 1),
\end{equation}
where $\mathcal{C}_2$ corresponds to the discrete symmetry related to $\tilde{S}$ and $\mathcal{C}_2'$ corresponds to the discrete symmetry related to $(\tilde{S}T)^2$. The Coulomb branch is not freely generated in any of these three cases, for $\mathbb{Z}_4$ it is not a complete intersection:
\begin{equation}
	\mathbb{Z}_4: \qquad \mathrm{PLog}\left\{\:\frac{1+U_2 U_6(U_2^2+U_2U_6+U_6^2)}{(1-U_2^4)(1-U^4_6)(1-U_8)(1-U_{12})}\right\},
\end{equation}
and in the two other cases the generators obey the relations 
\begin{align}
	&\mathbb{Z}_2: \qquad \tilde{u}_1 = U_2^2, \quad \tilde{u}_2 = U_6, \quad \tilde{u}_3 = U_8,\quad \tilde{u}_{4} = U_{12}^2,\quad \tilde{u}_5 = U_2 U_{12}; \quad \tilde{u}_5^2 = \tilde{u}_1\tilde{u}_4,\\
	&\mathbb{Z}_2': \qquad \tilde{u}_1 = U_2^2, \quad \tilde{u}_2 = U_6^2, \quad \tilde{u}_3 = U_8,\quad \tilde{u}_4 = U_{12},\quad \tilde{u}_{5} = U_2 U_6; \quad \tilde{u}_5^2 = \tilde{u}_1\tilde{u}_2.
\end{align}
The orbifolding group for the original manifold is $\mathrm{Weyl}(\mathfrak{f}_4) = G_{28}$, and its semidirect product with $\mathbb{Z}_2, \mathbb{Z}_4$ doesn't yield a complex reflection group.

The action of the generators $\mathcal{A}_l$ of the $\mathbb{Z}_l$ groups on the fields $\phi$ can be chosen to be
\begin{equation}
	\mathcal{A}_2 = i\cdot R, \qquad \mathcal{A}_4 = e^{i\pi/4} R, \qquad \mathcal{A}_2' = i\mathbb{1}; \qquad 	R = \frac{1}{\sqrt{2}}\begin{pmatrix}
		1 & -1 & 0 & 0 \\ 1 & 1 & 0 & 0 \\ 0 & 0 & 1 & -1 \\ 0 & 0 & 1 & 1
	\end{pmatrix}
\end{equation}
and the Higgs branch Hilbert series is given by
\begin{align}
	&\mathcal{I}_H^{F_4, \mathbb{Z}_2} = t^2 \left(\frac{u_2^2}{u_1}+u_1 u_2\right)+t^4 \left(u_1^4 v^4+\frac{u_2^4 v^4}{u_1^4}+\frac{u_2^3 v^4}{u_1^2}+\right.\\
	&\left.+2 u_2^2 v^4+u_1^2 u_2 v^4+\frac{u_2^4}{v^4}+\frac{u_2^4}{u_1^2}+u_2^3+u_1^2 u_2^2\right)+O(t^6)\\
	&\mathcal{I}_H^{F_4, \mathbb{Z}'_2} = t^2 \left(\frac{u_2^2}{u_1}+u_1 u_2\right)+t^4 \left(u_1^4 v^4+\frac{u_2^4 v^4}{u_1^4}+\right.+\frac{u_2^3 v^4}{u_1^2}\\
	&\left.+2 u_2^2 v^4+u_1^2 u_2 v^4+\frac{u_2^4}{v^4}+\frac{u_2^4}{u_1^2}+u_2^3+u_1^2 u_2^2\right)+O(t^6)\\
	&\mathcal{I}_H^{F_4, \mathbb{Z}_4} =  t^2 \left(\frac{u_2^2}{u_1}+u_1 u_2\right)+t^4 \left(\frac{u_2^4}{u_1^2}+u_2^3+u_1^2 u_2^2\right)+O(t^6)
\end{align}
$\mathcal{I}_H^{F_4, \mathbb{Z}_2}$ and $\mathcal{I}_H^{F_4, \mathbb{Z}'_2}$ differ at the $t^6$ order.

\subsubsection{Gauging \texorpdfstring{$N=4$ $B_n$}{N=4 Bn} SYM}
This case hasn't been analyzed in \cite{AKS}. As for the theories with $B_n$ (or $C_n$) gauge groups $S$ transformation takes us to another theory \cite{SeibergLines}, from the S-duality side we should only consider $\mathbb{Z}_2$ generated by $AT$. This transformation leaves the Coulomb branch invariant, so when we mix it with the R-symmetry part, we get that the action of the generator $\mathcal{C}_2$ of the $\Gamma_2$ groups on the Coulomb branch operators is given by
\begin{equation}
	\mathcal{C}_2 = \mathrm{diag}(-1, 1,\ldots, (-1)^n),
\end{equation}
where $n$ corresponds to the $B_n$ gauge group. For $B_2$ gauge group we get that the Coulomb branch is freely generated, the dimensions of Coulomb branch operators are $\Delta = 4, 4$, and the orbifolding group is $G(4, 2, 2)$, which is in agreement with \cite{Argyres1}.

For $B_3, B_4$ gauge groups we found that the Coulomb branch is not freely generated, and the generators obey the relations
\begin{align}
	& B_3: \qquad \tilde{u}_2 = U_2^2, \quad \tilde{u}_4 = U_4, \quad\tilde{u}_6 = U_6^2, \quad\tilde{u}_c = U_2 U_6; \qquad \tilde{u}_c^2 = \tilde{u}_2 \tilde{u}_6\\
	& B_4: \qquad \tilde{u}_2 = U_2^2,\quad\tilde{u}_4 = U_4, \quad\tilde{u}_6 = U_6^2, \quad\tilde{u}_8 = U_8,\quad \tilde{u}_c = U_2 U_6; \qquad \tilde{u}_c^2 = \tilde{u}_2 \tilde{u}_6
\end{align}
For $n\geq 5$ gauging $\Gamma_4$ yields a theory with Coulomb branch that is not a complete intersection manifold, the Molien series for e.g. $B_5$ is given by
\begin{equation}
	B_5: \qquad \frac{1+U_2 U_6 +U_2 U_{10}+U_6 U_{10}}{\left(1-U_2^2\right) \left(1-U_4\right) \left(1-U_6^2\right) \left(1-U_8\right) \left(1-U_{10}^2\right)}
\end{equation}

The action of the generators $\mathcal{A}_2$ of the $\mathbb{Z}_2$ groups on the fields $\phi$ can be chosen to be 
\begin{equation}
	\mathcal{A}_2 = i\mathbb{1}.
\end{equation}

The Higgs branch Hilbert series for $B_2 - B_5$ $N=4$ theories, gauged by $\mathbb{Z}_2$, at the two lowest orders is given by
\begin{align}
	&\mathcal{I}_H^{B} = t^2 \left(\frac{u_2^2}{u_1}+u_1 u_2\right)+t^4 \left(2 u_1^4 v^4+\frac{2 u_2^4 v^4}{u_1^4}+\frac{2 u_2^3 v^4}{u_1^2}+\right.\\
	&\left.+3 u_2^2 v^4+2 u_1^2 u_2 v^4+\frac{2 u_2^4}{v^4}+\frac{2 u_2^4}{u_1^2}+2 u_2^3+2 u_1^2 u_2^2\right)+O(t^6).
\end{align}
The difference between $B_2$ and $B_3$ indices appears at $t^6$ order, between $B_3$ and $B_4$ --- at $t^8$ order and between $B_4$ and $B_5$ --- at $t^{10}$ order.

\section{\texorpdfstring{$N=3$}{N=3} SCFTs from complex crystallographic reflection groups}
Another area of interest in $N=3$ SCFT studies is theories that have freely generated Coulomb branch. Until a few years ago it was commonly believed that every $N\geq 2$ theory possesses a freely generated Coulomb branch, but in \cite{Bourton, Argyres2} it was shown that there exist many $N=3$ SCFTs that possess a non-freely generated Coulomb branch. In \cite{Caorsi} Caorsi and Cecotti argued that for every $N=3$ SCFT with a freely generated Coulomb branch the Coulomb branch is given by $\mathbb{C}^r / G$, where $G$ is complex crystallographic reflection group (CCRG)\footnote{if $G$ is real, SUSY is enhanced to $N=4$.}. However, not every CCRG can give rise to a Coulomb branch of the $N=3$ SCFT; for example, a group $G_8$ cannot correspond to any $N=3$ SCFT \cite{Caorsi}. If a rank-$k$ CCRG $G$ corresponds to an $N=3$ SCFT, then consistency with Dirac quantization requires that $G \subset Sp(2k, \mathbb{Z})$\footnote{If this requirement is satisfied, then $G$ can be an orbifolding group for $N=3$ moduli space orbifold geometry; it doesn't mean there exist a corresponding $N=3$ SCFT.}. Caorsi and Cecotti argued that non-sporadic groups complex crystallographic reflection groups can always be embedded into $Sp(2k, \mathbb{Z})$, and in \cite{Argyres1} the issue of rank-2 sporadic groups has been addressed, so the groups left to analyze are $G_{24}$, $G_{25}$, $G_{26}$, $G_{29}$, $G_{31}$, $G_{32}$, $G_{33}$, $G_{34}$. In order to do that, we considered embedding of these groups into $GL(k, \mathbb{Z}[\zeta])$, where $k$ is the rank of the group and $\zeta$ is primitive third root of 1 for $G_{25}$, $G_{26}$, $G_{32}$, $G_{33}$, $G_{34}$; $\zeta = i$ for $G_{29}$, $G_{31}$ and $\zeta = \sqrt{-7}$ for $G_{24}$\footnote{One should also take into account that for $G_{25}$, $G_{26}$, $G_{33}$ there are two inequivalent embeddings, and both of them should be considered \cite{Weit}.}. Then we computed an invariant ($k \times k$) Hermitian form $H$ for each of these groups. Afterwards we constructed $2k \times 2k$ skew-symmetric form $\Omega$ from $H$ by considering 
\begin{equation}
	\frac{1}{\zeta - \bar{\zeta}}\;H_{ij}\; \psi^i \wedge \bar{\psi}^j,
\end{equation}
where $\psi^i = x^i + \zeta y^i \in \mathbb{Z}[\zeta]^k$. $\Omega$ is then obtained by clearing denominators and dividing by a non-trivial common factor for all the entries of the matrix in consideration if needed. Then the necessary and sufficient condition on whether the embedding $G \hookrightarrow GL(k, \mathbb{Z}[\zeta])$ induces an embedding $G \hookrightarrow Sp(2k, \mathbb{Z})$ is simply $\det \Omega = 1$. If there indeed exists an embedding $G \hookrightarrow Sp(2k, \mathbb{Z})$, then the complex crystallographic reflection group $G$ can correspond to an $N=3$ SCFT \cite{Argyres1}.

The direct computation shows that for $G =G_{24}$, $G_{25}$, $G_{26}$, $G_{32}$, $G_{33}$ $G\not\subset Sp(2k, \mathbb{Z})$. Let us list lowest terms of the Higgs branch Hilbert series expansion for $G_{29}$, $G_{31}$\footnote{$G_{34}$ has about $3.9\times 10^7$ elements, so it is computationally unfeasible to calculate Higgs branch Hilbert series for it.}:
\begin{align}
	& G_{29}: \qquad \mathcal{I}_H = t^2 \left(\frac{u_2^2}{u_1}+u_1 u_2\right)+t^4 \left(u_1^4 v^4+\right.\\
	&\left.+\frac{u_2^4 v^4}{u_1^4}+\frac{u_2^3 v^4}{u_1^2}+u_2^2 v^4+u_1^2 u_2 v^4+\frac{u_2^4}{v^4}\right)+O(t^6)\\
	& G_{31}: \qquad \mathcal{I}_H = t^2 \left(\frac{u_2^2}{u_1}+u_1 u_2\right)+O(t^8).
\end{align}
We can identify that at the $t^2$ order the only contribution to the index comes from the $N=3$ stress-tensor multiplet. The index expansion for $G_{31}$ has contributions at $t^2$ order and then only at $t^8$.

\bigskip
\noindent{\bf Acknowledgments}

I would like to thank Shlomo Razamat, Petr Kravchuk, and Andrey Feldman for useful discussions, and especially Ofer Aharony for useful discussions, general guidance and comments on a draft of this manuscript. This work was supported in part by an Israel Science Foundation center for excellence grant (grant number 1989/14) and by the Minerva foundation with funding from the Federal German Ministry for Education and Research.

\appendix

\section{Existence of theories listed in Bourton et al.} \label{sec:bourton}
In \cite{Bourton} Bourton et al. classify various $N=3$ SCFTs obtained from $N=4$ SCFTs with $ADE$ or $U(n)$ gauge groups. In particular, they have mixed finite cyclic subgroups of $SL(2,\mathbb{Z})$ self-duality groups with the $\mathbb{Z}_n \subset SU(R)_4$ of the R-symmetry group and then gauged the resulting group; for $n=3, 4, 6$ they have obtained $N=3$ SCFTs, while for $n=2$ they have got $N=4$ SCFTs. However, there is a fine point first observed in \cite{SeibergLines} related to the fact that there might be more than one theory for a given gauge algebra $\mathfrak{g}$, depending on the line operators present in the theory. $S$, $T$ transformations then may transform a theory with one set of line operators to a physically distinct theory with another set of line operators; an $N=4$ SCFT with gauge algebra $\mathfrak{g}$ (listed in \cite{Bourton}) will have a $Z_k \subset SL(2, \mathbb{Z})$ iff there is a theory which is self-dual under the corresponding $S$-duality transformation. Therefore, it turns out that not every $N=3$ SCFT listed in \cite{Bourton} exists; using \cite{SeibergLines} and \cite{Argyres2}, one can find that the following $N=3$ SCFTs exist:
\begin{center}
  \begin{tabular}{ | c | c | c | c| c|}
    \hline
      & $\mathbb{Z}_2$ & $\mathbb{Z}_3$ & $\mathbb{Z}_4$ & $\mathbb{Z}_6$ \\ \hline
    $SU(2)$ & + & $-$ & + & $-$ \\ \hline
    $SU(3)$ & + & + & $-$ & + \\ \hline
    $SU(4)$ & + & + & + & + \\ \hline
    $SU(5)$ & + & $-$ & + & $-$ \\ \hline
    $SO(2d)$, $d>1$ & + & + & + & + \\ \hline
    $U(d)$ & + & + & + & + \\ \hline    
    $E_6$ & + & + & $-$ & + \\ \hline
    $E_7$ & + & $-$ & + & $-$ \\ \hline
    $E_8$ & + & + & + & + \\ \hline
  \end{tabular}
\end{center}

These results are in agreement with \cite{Argyres1}, with the superficial exception of the $U(2)$ and $SO(4)$ $\mathbb{Z}_3$, $\mathbb{Z}_4$, $\mathbb{Z}_6$ gaugings. The classification of Argyres et al. contains the Coulomb branch orbifold geometries for these theories (Table 3, entry 32 for $\mathbb{Z}_3$ gauging of $U(2)$, Table 3, entry 33 for $\mathbb{Z}_4$ gauging of $U(2)$, $SO(4)$ and Table 4, entry 52 for $\mathbb{Z}_6$ gauging of $U(2)$ and  $\mathbb{Z}_3$, $\mathbb{Z}_6$ gauging of $SO(4)$). These entries were ruled out by Argyres et al. because the corresponding theories have two stress tensors (this can be seen from the Hilbert series analysis). However, as the mother $N=4$ theory had a non-simple gauge group, getting theory with two stress tensors after gauging a discrete subgroup is the expected outcome and does not mean the geometries in question should be discarded.


\begin{thebibliography}{99}
\bibitem{Bourton}
  T.~Bourton, A.~Pini and E.~Pomoni,
  ``4d $N=3$ indices via discrete gauging,''
  arXiv:1804.05396 [hep-th].
\bibitem{Argyres1}
  P.~C.~Argyres, A.~Bourget and M.~Martone,
  ``Classification of all $\mathcal{N}\geq 3$ moduli space orbifold geometries at rank 2,''
  arXiv:1904.10969 [hep-th].
\bibitem{Bonetti}
  F.~Bonetti, C.~Meneghelli and L.~Rastelli,
  ``VOAs labelled by complex reflection groups and 4d SCFTs,''
  JHEP {\bf 1905} (2019) 155
  doi:10.1007/JHEP05(2019)155
  [arXiv:1810.03612 [hep-th]].
\bibitem{Aharony}
  O.~Aharony and Y.~Tachikawa,
  ``S-folds and 4d N=3 superconformal field theories,''
  JHEP {\bf 1606} (2016) 044
  doi:10.1007/JHEP06(2016)044
  [arXiv:1602.08638 [hep-th]].
\bibitem{Shapere}
  A.~D.~Shapere and Y.~Tachikawa,
  ``Central charges of N=2 superconformal field theories in four dimensions,''
  JHEP {\bf 0809} (2008) 109
  doi:10.1088/1126-6708/2008/09/109
  [arXiv:0804.1957 [hep-th]].
\bibitem{Argyres2}
  P.~C.~Argyres and M.~Martone,
  ``Coulomb branches with complex singularities,''
  JHEP {\bf 1806} (2018) 045
  doi:10.1007/JHEP06(2018)045
  [arXiv:1804.03152 [hep-th]].
\bibitem{Argyres3}
  P.~C.~Argyres, C.~Long and M.~Martone,
  ``The Singularity Structure of Scale-Invariant Rank-2 Coulomb Branches,''
  JHEP {\bf 1805} (2018) 086
  doi:10.1007/JHEP05(2018)086
  [arXiv:1801.01122 [hep-th]].
\bibitem{Argyres4}
  P.~C.~Argyres and M.~Martone,
  ``4d $ \mathcal{N} $ =2 theories with disconnected gauge groups,''
  JHEP {\bf 1703} (2017) 145
  doi:10.1007/JHEP03(2017)145
  [arXiv:1611.08602 [hep-th]].
\bibitem{Argyres5}
  P.~Argyres, M.~Lotito, Y.~Lü and M.~Martone,
  ``Geometric constraints on the space of $ \mathcal{N}$ = 2 SCFTs. Part III: enhanced Coulomb branches and central charges,''
  JHEP {\bf 1802} (2018) 003
  doi:10.1007/JHEP02(2018)003
  [arXiv:1609.04404 [hep-th]].
\bibitem{Caorsi}
  M.~Caorsi and S.~Cecotti,
  ``Geometric classification of 4d $\mathcal{N}=2$ SCFTs,''
  JHEP {\bf 1807} (2018) 138
  doi:10.1007/JHEP07(2018)138
  [arXiv:1801.04542 [hep-th]].
\bibitem{Todd}
 Shephard,~G., and Todd,~J. 
 (1954). Finite Unitary Reflection Groups. 
 Canadian Journal of Mathematics, 6, 274-304. 
 doi:10.4153/CJM-1954-028-3
\bibitem{SeibergLines}
  O.~Aharony, N.~Seiberg and Y.~Tachikawa,
  ``Reading between the lines of four-dimensional gauge theories,''
  JHEP {\bf 1308} (2013) 115
  doi:10.1007/JHEP08(2013)115
  [arXiv:1305.0318 [hep-th]].
\bibitem{AKS}
  P.~C.~Argyres, A.~Kapustin and N.~Seiberg,
  ``On S-duality for non-simply laced gauge groups,''
  JHEP {\bf 0606} (2006) 043
  doi:10.1088/1126-6708/2006/06/043
  [hep-th/0603048].
\bibitem{Kapustin}
  A.~Kapustin and E.~Witten,
  ``Electric-Magnetic Duality And The Geometric Langlands Program,''
  Commun.\ Num.\ Theor.\ Phys.\  {\bf 1} (2007) 1
  doi:10.4310/CNTP.2007.v1.n1.a1
  [hep-th/0604151].
\bibitem{Weit}
  W.~Feit, 
  ``Some integral representation of complex reflection groups,'' 
  J. of Algebra {\bf 260} (2003) 138-153
  doi:10.1016/S0021-8693(02)00629-4
(2003) 138-153.
\end{thebibliography}
\end{document}